\documentclass[journal]{IEEEtran}
\usepackage{amsmath,,amssymb,amsfonts}
\usepackage{algorithmic}
\usepackage{algorithm}
\usepackage{array}
\usepackage{color}
\usepackage[caption=false,font=normalsize,labelfont=sf,textfont=sf]{subfig}
\usepackage{textcomp}
\usepackage{stfloats}
\usepackage{url}
\usepackage{bm}
\usepackage{verbatim}
\usepackage{graphicx}
\usepackage{cite}

\usepackage{amsthm}
\usepackage{caption}
\usepackage{graphicx}
\usepackage{float} 
\usepackage{subcaption}
\begin{document}

\title{Joint Admission Control and Power Minimization in IRS-assisted Networks}

\author{Weijie Xiong, Jingran Lin, Zhiling Xiao, Qiang Li, and Yuhan Zhang
\thanks{W. Xiong, J. Lin, Z. Xiao, Q. Li, and Y. Zhang are with the School of Information and Communication Engineering, University of Electronic Science and Technology of China, Chengdu 611731, China. J. Lin and Q. Li are also with the Laboratory of Electromagnetic Space Cognition and Intelligent Control, Beijing 100083, China, and the Tianfu Jiangxi Laboratory, Chengdu, Sichuan 641419, China. }
\thanks{This work was supported in part by the Natural Science Foundation of China (NSFC) under Grant 62171110.  \textit{(Corresponding author: Jingran Lin, e-mail: jingranlin@uestc.edu.cn)}.}
}

\markboth{Journal of \LaTeX\ Class Files,~Vol.~14, No.~8, August~2021}%
{Shell \MakeLowercase{\textit{et al.}}: A Sample Article Using IEEEtran.cls for IEEE Journals}


\maketitle

\begin{abstract}
Joint admission control and power minimization are critical challenges in intelligent reflecting surface (IRS)-assisted networks. Traditional methods often rely on \( l_1 \)-norm approximations and alternating optimization (AO) techniques, which suffer from high computational complexity and lack robust convergence guarantees. To address these limitations, we propose a sigmoid-based approximation of the \( l_0 \)-norm AC indicator, enabling a more efficient and tractable reformulation of the problem. Additionally, we introduce a penalty dual decomposition (PDD) algorithm to jointly optimize beamforming and admission control, ensuring convergence to a stationary solution. This approach reduces computational complexity and supports distributed implementation. Moreover, it outperforms existing methods by achieving lower power consumption, accommodating more users, and reducing computational time.
\end{abstract}


\section{Introduction}
Power minimization is a critical aspect of intelligent reflecting surface (IRS)-assisted systems. By jointly optimizing transmit beamforming and IRS phase shifts, these systems enhance signal quality, mitigate interference, and significantly reduce the base station's power consumption while meeting quality-of-service (QoS) requirements \cite{zhou2020framework,zhou2022channel,mitev2023physical}. However, the increasing number of users has made it challenging to satisfy QoS demands for all users simultaneously, often resulting in network collapse \cite{lin2020joint,bereyhi2021securing,khisti2010secure}.

To address this challenge, admission control (AC) is widely employed to selectively reject a minimal number of access requests while ensuring satisfactory QoS for the remaining users \cite{liu2020simple,sato2022riemannian,ring2012optimization}. Most AC strategies involve solving complex integer-mixed programming problems due to the \( l_0 \)-norm AC indicator. To balance performance and computational complexity, methods such as semidefinite relaxation (SDR) \cite{matskani2008convex}, second-order cone programming (SOCP) \cite{lin2020joint}, and linear programming deflation \cite{mitliagkas2011joint} are commonly used to obtain suboptimal solutions.

However, applying AC to IRS-assisted networks presents additional challenges due to the complexity of managing the highly coupled variables between the IRS and transmit beamforming. This coupling results in a high-dimensional, non-convex optimization landscape, making efficient solutions difficult. Existing methods often use \(l_1\)-norm approximations and alternating optimization (AO) algorithms \cite{li2021joint,he2023reconfigurable}, but they lack theoretical convergence guarantees and rely on the interior point method (IPM) for subproblems, leading to high computational complexity for large variable dimensions \cite{lin2020joint}.

To address these challenges, this paper makes two key contributions to enhance efficiency and convergence. Firstly, we develop a continuous function to approximate the \( l_0 \)-norm AC indicator, making the problem more tractable. Unlike conventional \( l_1 \)-norm approximations, this sigmoid-based approach more accurately captures the \( l_0 \)-norm. Secondly, we propose a penalty dual decomposition (PDD) algorithm for joint beamforming and admission control, ensuring stationary convergence while reducing computational complexity and enabling efficient distributed implementation. Compared to the AO methods in \cite{li2021joint, he2023reconfigurable}, simulation results highlight its advantages: robust convergence regardless of hyperparameter selection, 0.9 and 0.4 more admitted users, 0.12W and 0.06W lower power consumption, and convergence times 10x and 100x faster.

\section{System Model and Problem Formulation}
Consider an IRS-assisted downlink cellular network, where an $N$-antenna base station (BS) serves $M$ single-antenna users with the assistance of an IRS. The IRS consists of $K$ passive reflecting elements, each capable of independently adjusting its phase shift. The channel coefficients of the BS-IRS link, the IRS-user $m$ link, and the BS-user $m$ link are denoted as ${\bf G} \in {\mathbb C}^{K \times N}$, ${\bf h}_m \in {\mathbb C}^{K \times 1}$, and ${\bf g}_m \in {\mathbb C}^{N \times 1}$, respectively. Let ${\bf w}_m \in {\mathbb C}^{N \times 1}$ be the BS transmit beamformer for user $m$\footnote{In our analysis, unlike works that consider imperfect \cite{zhou2020framework} or cascaded \cite{zhou2022channel} CSI, we assume the ideal case where perfect CSI is available. This assumption is feasible with advanced estimation methods, particularly in low-mobility environments where channels change slowly \cite{he2023reconfigurable}, and by leveraging recent advancements in deep learning and pilot training schemes \cite{chen2024coded}.}. The signal-to-interference-plus-noise ratio (SINR) is used to evaluate user QoS, and the SINR of user $m$ is,
\begin{equation}
\begin{aligned}
   {\text {SINR}}_m&=\tfrac{|({\bf g}_m^H+{\bf h}_m^H {\bf \Theta} {\bf G}) {\bf w}_m|^2}{\sum_{n \neq m} |({\bf g}_m^H+{\bf h}_m^H {\bf \Theta} {\bf G}) {\bf w}_n|^2+\sigma_m^2} \\
   &=\tfrac{|{\bf p}_m({\bm \theta}) {\bf w}_m|^2}{\sum_{n \neq m} |{\bf p}_m({\bm \theta}) {\bf w}_n|^2+\sigma_m^2},\;\forall m, 
\end{aligned}
\end{equation}
where ${\bf h}_m^H {\bf \Theta} {\bf G} = {\bm \theta}^T{\text {Diag}}({\bf h}_m^H){\bf G}$; ${\bf q}_m   = {\text {Diag}}({\bf h}_m^H){\bf G}$; ${\bf p}_m({\bm \theta})  = {\bf g}_m^H+{\bm \theta}^T{\bf q}_m$; ${\bf \Theta} = {\text {Diag}}({\bm \theta}) \in {\mathbb C}^{K \times K}$ denotes the diagonal phase-shift matrix of the IRS; ${\bm \theta} = [\theta_1,\theta_2,...,\theta_K]^T \in \mathbb{C}^{K}$ is the reflecting coefficients at the IRS; $\sigma_m$ is the floor noise power at user $m$; $(\cdot)^H$ and $(\cdot)^T$ denote Hermitian transpose and transpose, repectively.

In traditional power minimization problems, the BS transmit beamformer and the IRS phase shifts are jointly optimized to minimize the transmit power. Considering the SINR constraints for the users and the maximum transmit power constraint of the BS, the problem is formulated as,
\begin{subequations}
\begin{align}
 \min _{\{{\bf W}, {\bm \theta}\}} & \|{\bf W}\|_F^2, \\
 {\text {s.t.}} \quad &{\text {SINR}}_m  \ge {\gamma},\; \forall m, \\
 \quad\quad &\|{\bf W}\|_F^2  \le P, \\
 \quad\quad &|{\theta}_k | = 1, \; \forall k,
\end{align}
\end{subequations}
where ${\bf W} = [{\bf w}_1,{\bf w}_2,...,{\bf w}_M] \in {\mathbb C}^{N \times M}$ denotes the beamformer matrix; ${\gamma}$ is the QoS threshold; $P$ is the BS transmit power budget; $||\cdot||_F$ denotes the Frobenius norm. 

Note that (2b) may not always be feasible due to the surge in user numbers and the resulting intense competition. Consequently, AC becomes indispensable. Specifically, to derive a feasible solution to problem (2), an auxiliary variable ${\bf a} = [a_1; a_2; \dots; a_M] \in {\mathbb C}^{M }$ is typically introduced into problem (2), and the SINR constraint in (2b) is reformulated using second-order cone programming (SOCP) as \cite{lin2020joint},
\begin{equation}
\left\{\begin{array}{l}
{\bf p}_m({\bm \theta}) {\bf w}_m + a_m \geq \sqrt{\gamma_m (\sigma_m^2+\sum_{n \neq m}\left|{\bf p}_m({\bm \theta}) {\bf w}_n\right|^2) }, \\
\Im\left\{{\bf p}_m({\bm \theta}) {\bf w}_m\right\}=0, \; \forall m,
\end{array}\right.
\end{equation} 
where $a_m$ is the gap of received signal power for user $m$ to satisfy its desired QoS requirement; $\Im\{\cdot\}$ denotes the imaginary part of the argument. 

Moreover, we aim to reject as few users as possible while meeting the QoS requirements of the remaining users. To achieve this, an $l_0$-norm AC indicator $\|{\bf a}\|_0$ is introduced in the objective function as a penalty term, and the joint design problem is formulated as,
\begin{subequations}
\begin{align}
 \min _{\{{\bf W}, {\bm \theta}, {\bf a}\}} & \|{\bf W}\|_F^2 +  \lambda \|{\bf a}\|_0, \\
 {\text {s.t.}} \quad &a_m \geq 0,\;\forall m,\\
 \quad\quad&{\text {(2c), (2d), (3) are satisfied.}} 
\end{align}
\end{subequations}
where $a_m = 0$ indicates that the QoS level of user $m$ can be achieved and hence user $m$ is admissible, while $a_m > 0$ indicates that user $m$ is inadmissible; $\lambda>0$ is a large positive constant to balances the network power cost and the size of admissible user set. In practice, \(\lambda\) serves as a tunable hyperparameter to adapt to specific system requirements. Benefiting from the sparsity induced by $\|{\bf a}\|_0$, the penalty term in the objective function results in two binary outcomes, minimizing the number of users denied.

Problem (4) is non-convex due to the non-continuity $l_0$-norm AC indicators $\|{\bf a}\|_0$ in (4a), the highly-coupled non-linear variables $\{{\bf W},{\bm \theta} \}$ in (3) and the constant modulus constraints (CMC) in (2d). In light of the above difficulties, we focus on developing approaches to find high-quality approximate solutions to problem (4) in the following sections.

\section{The PDD-based Algorithm}
In general, the \(l_0\)-norm AC indicator \(\|\mathbf{a}\|_0\) is non-continuous and challenging to solve directly. Existing AC methods often approximate the \(l_0\)-norm with the \(l_1\)-norm \cite{matskani2008convex,li2021joint}. Although efficient, this approach may result in significant information loss, particularly in IRS-assisted communication networks, where the strong coupling between variables reduces the effectiveness of conventional \(l_1\)-relaxation methods. Recent studies suggest that sigmoid functions, widely used in machine learning, provide a better approximation of the \(l_0\)-norm \cite{kyurkchiev2015sigmoid}. Inspired by these findings, we adopt a sigmoid function to approximate the AC indicator, 
\begin{equation}
\|{\bf a}\|_0  \approx \sum_{m=1}^{M} { \mathcal{I}(a_m)} \triangleq \sum_{m=1}^{M} (1-\frac{1}{\text{exp}({\Gamma a_m})}), a_m \geq 0,
\end{equation}
where \(\exp(\cdot)\) is the exponential function, and \(\Gamma\) is a sufficiently large integer. When \(a_m = 0\), the function value is 0, and when \(a_m > 0\), it quickly approximates to 1. This approach retains more information and better represents the original \( l_0 \)-norm, making it well-suited for managing the strong coupling between variables in IRS-assisted systems. The interdependence of the transmit beamformer and IRS phase shifts poses challenges for traditional \( l_1 \)-relaxation methods, which this approach effectively overcomes by enabling more accurate optimization.

Then, to circumvent the non-linear constraint (3), we introduce a series of variables ${\bm \Psi} = [{\bm \psi}_1, {\bm \psi}_2,..., {\bm \psi}_M] \in {\mathbb C}^{K \times M}$ and ${\bf E} \in {\mathbb C}^{M \times (M+1)}$. Define the $m$th row of ${\bf E}$ as ${\bf e}^m = [e_1^m,e_2^m,...,e_M^m,e_{M+1}^m]\in{\mathbb C}^{1 \times (M+1)}$. In the case of ${\bm \psi}_1 = {\bm \psi}_2 = ...={\bm \psi}_M={\bm \theta}$; $e_n^m = {\bf p}_m({\bm \psi}_m) {\bf w}_n,n=1,...,M$, $e_{M+1}^m = 1$; ${\bf a}={\bf c}$, problem (5) can be equivalently recast as,
\begin{subequations}
\begin{align}
 \min _{\{{\bf W}, {\bm \theta}, {\bf a},{\bm \Psi}, {\bf E},{\bf c}\}} & \|{\bf W}\|_F^2 +   \lambda \textstyle\sum_{m=1}^M \mathcal{I}(c_m), \\
 \text{s.t.} \quad\quad &e_m^m + a_m \ge \sqrt{\gamma_m} \cdot  \| {\bf e}_{-m}^m \|_2,\; \Im\{e_m^m\}=0,\; \forall m,\\
 \quad\quad &\|{\bf W}\|_F^2  \le P, \\ 
 \quad\quad &|{\theta}_k | = 1, \; \forall k, \\
 \quad\quad &c_m \geq 0,\;\forall m,\\
 \quad\quad &{\bm \psi}_1 = {\bm \psi}_2 = ...={\bm \psi}_M={\bm \theta},\\
 \quad\quad &{\bf E} = [{\bf P}({\bf \Psi}){\bf W}, \bm{\sigma}],\\
 \quad\quad &{\bf c}={\bf a}.
\end{align}
\end{subequations}
Notice that ${\bf P}({\bf \Psi}) = \left[{\bf p}_1({\bm \psi}_1); {\bf p}_2({\bm \psi}_2);...;{\bf p}_M({\bm \psi}_M) \right] \in {\mathbb{C}^{M\times N} }$ where ${\bf p}_m({\bm \psi}_m)= {\bf g}_m^H+{\bm \psi}_m^T{\bf q}_m, \forall m $; ${\bf e}_{-m}^m=[e_1^m,...,e_{m-1}^m,e_{m+1}^m,...,e_{M+1}^m]\in{\mathbb C}^{1 \times M}$ is obtained by removing $e_m^m$ from ${\bf e}^m$; $\bm{\sigma} = [{\sigma}^2_1,{\sigma}^2_2,...,{\sigma}^2_M] \in {\mathbb{C}^{M\times 1} }$.

In problem (6), the challenging SINR constraint (3) is separated
into two parts: (i) the convex SOCP SINR constraint (6b), and (ii) the
non-convex and (${\bf P}({\bf \Psi})$,$\bf W$)-coupling equality constraint (6g). Obviously, the latter prevents the utilization of block decomposition methods. This motivates us to consider the method of PDD \cite{shi2020penalty}. To decouple variables, we dualize equality constraints by proper penalty functions, and thus ${\{{\bf W}, {\bm \theta}, {\bf a},{\bm \Psi}, {\bf E}, {\bf c}\}}$ can be updated iteratively following the coordinate descent framework.  

\subsection{Framework of the PDD-based Algorithm}
The equality constraint dualization is performed via Lagrangian relaxation, incorporating equality constraints into the objective function with Lagrange multipliers to simplify the optimization. This transforms constraints into penalty terms, balancing objective minimization with constraint satisfaction. The augmented Lagrangian (AL) function guides optimization towards feasible solutions while treating objectives and constraints in a unified manner. The AL function is defined as,
\begin{equation}
\begin{aligned}
&{\cal L}_{\rho} ({\bf W}, {\bm \theta}, {\bf a},{\bm \Psi}, {\bf E}, {\bf c}; {\bm \Xi}, {\bm \Phi}, {\bm \zeta } ) = \|{\bf W}\|_F^2 +  \lambda \textstyle\sum_{m=1}^M \mathcal{I}(c_m) \\& + \textstyle\sum_{m=1}^M [ \Re \{  {\bm \xi}_m^H ( {\bm \psi}_m - {\bm \theta} ) \}  +  \tfrac{1}{2\rho} \| {\bm \psi}_m - {\bm \theta} \|_2^2 ] \\&+  \Re \{ \text{Tr}\{ {\bm \Phi}^H ( {\bf E} - [{\bf P}({\bm \Psi}){\bf W},\bm{\sigma}] ) \} \} \\& + \tfrac{1}{2\rho} \|  {\bf E} - [{\bf P}({\bm \Psi}){\bf W},\bm{\sigma}]  \|_F^2 \\& + \Re \{  {\bm \zeta}^H ( {\bf c} - {\bf a} ) \} + \tfrac{1}{2\rho} \| ( {\bf c} - {\bf a} ) \|_2^2,
\end{aligned}
\end{equation}
where ${\bm \Xi} = [{\bm \xi}_1,{\bm \xi}_2,...,{\bm \xi}_M]\in{\mathbb C}^{K \times M}$, ${\bm \Phi} \in{\mathbb C}^{K \times M}$ and ${\bm \zeta} \in{\mathbb C}^{M \times 1}$ are the Lagrangian multipliers with (6f), (6g) and (6h), respectively; ${\rho}>0$ is the penalty parameter; $\Re\{\cdot\}$ denotes the real part of the argument; $\text{Tr}\{\cdot\}$ computes the trace of a matrix. Based on (7), we construct the AL problem below, providing a clear optimization framework for the subsequent analysis,
\begin{equation}
\begin{aligned}
 (\mathcal{P}_{\mathcal{L}_{\rho}} ): \min _{\{{\bf W}, {\bm \theta}, {\bf a}, {\bm \Psi}, {\bf E}, {\bf c}\}} &\mathcal{L}_{\rho}({\bf W}, {\bm \theta}, {\bf a},{\bm \Psi}, {\bf E}, {\bf c}; {\bm \Xi}, {\bm \Phi}, {\bm \zeta}), \\
 \text{s.t.} \quad\quad &\text { (6b), (6c), (6d), (6e) are satisfied. } \\
\end{aligned}
\end{equation}

According to the PDD framework, the PDD-based approach to problem (8) is summarized in Algorithm 1. Here, '$\text{OPT}(\mathcal{P}_{\mathcal{L}{\rho}}; \vartheta)$' represents the update of $\{{\bf W},  {\bm \theta}, {\bf a}, {\bm \Psi}, {\bf E}, {\bf c} \}$ by solving the problem $\mathcal{P}_{\mathcal{L}{\rho}}$ using a specific algorithm until the threshold $\vartheta$ is reached. The parameter $\sigma$ represents the maximum violation of the equality constraints. If $\sigma \le \eta$, Algorithm 1 updates the Lagrangian multipliers using the dual ascent method; otherwise, it adjusts the penalty parameter to enforce the relaxed equality constraints. Algorithm 1 stops once $\sigma \le \tau$. We set $b_1$ and $b_2$ within the interval $(0, 1)$ to ensure convergence, which are employed to iteratively decrease $\rho$, $\eta$, and $\vartheta$.
\begin{algorithm}
	\floatname{algorithm}{Algorithm}
	\renewcommand{\algorithmicrequire}{\textbf{Input:}}
	\renewcommand{\algorithmicensure}{\textbf{Output:}}
	\caption{: The PDD-based algorithm to the problem (9).}
	\begin{algorithmic}[1]
		\REQUIRE 
                ${\bf W}, {\bm \theta}, {\bf a}, {\bm \Psi}, {\bf E}, {\bm \Xi}, {\bm \Phi}, \rho, \sigma , \vartheta, \eta,\tau, b_1,$ \text{and} $b_2$.\\
            \STATE {\textbf {Reapeat}}
            \STATE  $\{{\bf W}, {\bm \theta}, {\bf a}, {\bm \Psi}, {\bf E}, {\bf c} \} = \text{OPT}(\mathcal{P}_{\mathcal{L}_{\rho}}; \vartheta )$;
            \STATE 
            $ \sigma = \text{max}\{\|{\bm \psi}_m - {\bm \theta}\|_\infty, \forall m, \; \|{\bf E} - [{\bf P}({\bf \Psi}){\bf W}, {\bm \sigma}]\|_\infty ,\|{\bf a} - {\bf c}\|_\infty\} $;
            \STATE  \textbf{IF} $\sigma <\eta $
            \STATE \quad ${\bm \xi}_m ={\bm \xi}_m +{\rho^{-1}}({\bf v}-{\bf{w}}),\; \forall m$;
            \STATE \quad ${\bm \Phi} ={\bm \Phi} +{\rho^{-1}}( {\bf E} - [{\bf P}({\bf \Psi}){\bf W}, \bm{\sigma}])$;
             \STATE \quad ${\bm \zeta} ={\bm \zeta} +{\rho^{-1}}({\bf a}-{\bf{c}})$;
            \STATE  \textbf{ELSE}
            \STATE \quad ${\rho = b_1 \cdot \rho }$;
            \STATE  \textbf{END IF}
            \STATE ${\eta = b_2 \cdot \sigma }$, and ${\vartheta = b_2 \cdot \vartheta }$
		\STATE {\textbf {Until } stopping criterion (e.g. $\sigma  \le \tau $)  satisfied.}\\
	\end{algorithmic}%
\end{algorithm}

\subsection{The BSUM-Based Algorithm for AL Problem}
The primary complexity of Algorithm 1 arises in step 2, which involves solving the problem $\mathcal{P}_{\mathcal{L}{\rho}}$ to update the variables $\{{\bf W}, {\bm \theta}, {\bf a}, {\bm \Psi}, {\bf E}, {\bf c}\}$. Although the problem ${\mathcal P}{\mathcal{L}{\rho}}$ is non-convex, its separable constraints enable block decomposition into five subproblems related to ${{\bf W}}$, ${{\bm \theta}}$, ${{\bm \Psi}}$, ${{\bf c}}$, and ${{\bf a}, {\bf E}}$. This approach facilitates distributed implementation and enhances convergence by cyclically solving the subproblems. To address the non-convexity of certain subproblems, we employ the BSUM method \cite{razaviyayn2013unified}, which constructs locally tight upper bounds to transform them into convex problems. We next demonstrate the application of BSUM to solve $\mathcal{P}{\mathcal{L}{\rho}}$.
\subsubsection{Update $\bf W$}
With ${\bm \theta}$, ${\bf a}$, ${\bm \Psi}$ and ${\bf E}$ being fixed, the subproblem related to $\bf W$ is convex and expressed as,
\begin{equation}
\begin{aligned}
& \min _{{\bf W}}\|\mathbf{W}\|_F^2 -  \Re \{ \text{Tr}\{ {\bm \Phi}_M^H {\bf P}({\bm \Psi}){\bf W} \} \} \\&\quad\quad\quad\quad\quad\quad\quad\quad\quad\quad\quad+ \tfrac{1}{2\rho} \| ( {\bf E}_M - {\bf P}({\bm \Psi}){\bf W} ) \|_F^2, \\
& \text { s.t. }\|\mathbf{W}\|_F^2 \leq P,
\end{aligned}
\end{equation}
where ${\bm \Phi}_M$ and ${\bf E}_M$ are the $M \times M$ left submatrices of ${\bm \Phi}$ and ${\bf E} \in \mathbb{C}^{M \times {M+1}}$, respectively. By checking the first-order optimality condition, the optimal $\bf W$ is given by,
\begin{equation}
{\bf W} = [  {\bf P}({\bm \Psi})^H{\bf P}({\bm \Psi}) + 2 \rho (1 + \alpha) {\bf I}  ]^{-1}{\bf P}({\bm \Psi})^H(\rho{\bm \Phi}_M+{\bf E}_M),
\end{equation}
where $\alpha$ is the Lagrangian multiplier chosen to satisfy the Karush-Kuhn-Tucker (KKT) conditions. To this end, we rewrite ${\bf P}({\bm \Psi})^H{\bf P}({\bm \Psi})$ in the eigen-decomposition form, i.e.,
\begin{equation}
{\bf P}({\bm \Psi})^H{\bf P}({\bm \Psi}) = {\bf U} {\bm \Pi} {\bf U}^H,
\end{equation}
where ${\bm \Pi} = \text{Diag}\{{\bm \pi}_1,{\bm \pi}_2,...,{\bm \pi}_N  \}$ with ${\bm \pi}_n$ being the $n$th nonnegative eigenvalue of ${\bf P}({\bm \Psi})^H{\bf P}({\bm \Psi})$; ${\bf U}=[ {\bf u}_1,{\bf u}_2,...,{\bf u}_N ]$ is the unitary matrix with ${\bf u}_n \in \mathbb{C}^{N \times 1}$ is the $n$th eigenvector of ${\bf P}({\bm \Psi})^H{\bf P}({\bm \Psi})$. We further define ${\bf \Delta} \buildrel \Delta \over = {\bf U}^H{\bf P}({\bm \Psi})^H( \rho{\bm \Phi}_M+{\bf E}_M )( \rho{\bm \Phi}_M+{\bf E}_M )^H {\bf P}({\bm \Psi}){\bf U}$, and have,
\begin{equation}
\|\mathbf{W}\|_F^2 = \text{Tr}\{ \mathbf{W} \mathbf{W}^H \} = \textstyle\sum_{n=1}^N \tfrac{\delta_n^n }{[\pi_n+2\rho(1+\alpha)]^2},
\end{equation}
with $\delta_n^n$ being the $n$th diagonal element of ${\bf \Delta}$. Therefore, in the case of $\sum_{n=1}^N \tfrac{\delta_n^n }{(\pi_n+2\rho)^2} \le P$, we have $\alpha = 0$; otherwise, we find certain $\alpha > 0$ so that (12) holds for equality, which can be done by bisection, since $\|\mathbf{W}\|_F^2$ is monotonically decreasing with $\alpha$.\footnote{To compute \(\mathbf{W}\) by (12) via the bisection method, first evaluate \(S_0 = \sum_{n=1}^N \frac{\delta_n^n}{(\pi_n + 2\rho)^2}\). If \(S_0 \leq P\), set \(\alpha = 0\) as the power constraint is satisfied. Otherwise, initialize the bisection interval with \(\alpha_{\text{low}} = 0\) and choose an upper bound \(\alpha_{\text{high}}\) such that \(\sum_{n=1}^N \frac{\delta_n^n}{[\pi_n + 2\rho(1 + \alpha_{\text{high}})]^2} < P\). Iteratively compute the midpoint \(\alpha_{\text{mid}} = (\alpha_{\text{low}} + \alpha_{\text{high}})/2\) and evaluate \(S_{\text{mid}} = \sum_{n=1}^N \frac{\delta_n^n}{[\pi_n + 2\rho(1 + \alpha_{\text{mid}})]^2}\). If \(S_{\text{mid}} > P\), update \(\alpha_{\text{low}} = \alpha_{\text{mid}}\); otherwise, set \(\alpha_{\text{high}} = \alpha_{\text{mid}}\). Repeat this process until \(|S_{\text{mid}} - P|\) is within a predefined tolerance. Once the optimal \(\alpha^*\) is determined, substitute it back into equation (12) to obtain \(\mathbf{W}\), thereby ensuring that \(\|\mathbf{W}\|_F^2 = P\).}

\subsubsection{Update $\bm \theta$}
With ${\bf W}$, ${\bf a}$, ${\bm \Psi}$ and ${\bf E}$  being fixed, the subproblem related to $\bm \theta$ is equivalent to,
\begin{equation}
\begin{aligned}
& \min _{{\bm \theta}} \textstyle\sum_{m=1}^M \tfrac{1}{2\rho} \|  {\bm \theta}- ({\bm \psi}_m  + \rho{\bm \xi}_m) \|_2^2, \\
& \text { s.t. }|{\theta}_k| = 1, \; \forall k,
\end{aligned}
\end{equation}
which admits the following closed-form solution, following equations (13) and (14) in \cite{9478276},
\begin{equation}
{\bm \theta} = \exp\left(j,\text{arg}\left(\frac{1}{M} \sum_{m=1}^M ({\bm \psi}_m + \rho{\bm \xi}_m)\right)\right), 
\end{equation}
where $\text{arg}(\cdot)$ denotes the argument of a complex vector. Note that Equation (14) provides the optimal solution to subproblem (13), although the CMC related to $\bm{\theta}$ is generally non-convex, as indicated by Lemma VI.1 in \cite{9478276}.

\subsubsection{Update ${\bm \Psi}$}
It is easy to see that the subproblem with respect to ${\bm \Psi}$ takes a separable structure, and can be decomposed into $M$ smaller problems related to ${\bm \psi}_m$, $m=1,2,...,M$, respectively. The individual problem of ${\bm \psi}_m$ is expressed as,
\begin{equation}
\begin{aligned}
& \min _{{\bm \psi}_m} \Re \{  {\bm \xi}_m^H \left( {\bm \psi}_m - {\bm \theta} \right) \}  + \tfrac{1}{2\rho} \| {\bm \psi}_m - {\bm \theta} \|_2^2  \\&+  \Re \{ \text{Tr}\{  ({{\bm \phi}_M^m})^H ( {\bf e}_M^m - {\bf p}_m({\bm \psi}_m){\bf W} ) \} \}\\&+ \tfrac{1}{2\rho} \| ( {\bf e}_M^m - {\bf p}_m({\bm \psi}_m){\bf W}  ) \|_2^2,   
\end{aligned}
\end{equation}
where ${\bm \phi}_M^m$ and ${\bf e}_M^m$ are the $1 \times M$ left subrows of the $m$th row of ${\bm \Phi}$ and ${\bf E}$, respectively. It is an unconstrained convex problem, and the optimal ${\bm \psi}_m$ is given as,
\begin{equation}
\begin{aligned}
{\bm \psi}_m = &[({\bf q}_m{\bf W})^H({\bf q}_m{\bf W}) +  \bf{I} ]^{-1} \cdot \\& ( ({\bf q}_m{\bf W}) ( 
 \rho{\bm \phi}_M^m + ({\bf g}_m^H{\bf W} - {\bf e}_M^m )^H)  - \rho {\bm \xi}_m + {\bm \theta} ).
\end{aligned}
\end{equation}

\subsubsection{Update ${\bf c}$}
The resultant subproblem related to ${\bf c}$ can be decomposed into $M$ smaller problems with respect to ${c_m,m=1,2,...,M}$. The subproblem of ${c_m}$ is given as, 
\begin{equation}
\begin{aligned}
& \min _{{ c_m}} \lambda \mathcal{I}(c_m) + \zeta_m c_m + \tfrac{1}{\rho}(c_m-a_m)^2, \\
& \text { s.t. }c_m \ge 0 .
\end{aligned}
\end{equation}
Problem (17) is non-convex due to the concave function \(\mathcal{I}(\cdot)\). To address this non-convexity, we apply the BSUM method by finding a locally tight upper bound. Specifically, by employing the first-order Taylor expansion at $\hat c_m$, we have \cite{lin2020joint},
\begin{equation}
\begin{aligned}
\mathcal{I}(c_m) = 1-\tfrac{1}{\text{exp}({\Gamma c_m})}  \le 1 - \tfrac{1}{\text{exp}({\Gamma {\hat c_m}})} + \tfrac{\Gamma( c_m- {\hat c_m})}{\text{exp}(\Gamma {\hat{c_m}})},
\end{aligned}
\end{equation}
where ${\hat{c_m}}$ is the value of $c_m$ in the previous iteration. Then, after ignoring the constant term, (18) is reformulated as,
\begin{equation}
\begin{aligned}
& \min _{{ c_m}} {\hat \lambda} c_m + \zeta_m c_m + \tfrac{1}{2\rho}(c_m-a_m)^2, \\
& \text { s.t. }c_m \ge 0,
\end{aligned}
\end{equation}
where ${\hat \lambda} = \tfrac{\Gamma \lambda}{\text{exp}(\Gamma {\hat{c_m}})}$. By checking the first-order optimality condition, the optimal $c_m$ is given by,
\begin{equation}
c_m = [ a_m - \rho ( {\hat \lambda} + \zeta_m  )]^{+},
\end{equation}
where $[\cdot]^+=\text{max}\{0,\cdot\}$.

\subsubsection{Update ${\bf E}$ and $\bf a$}
Update ${\bf E}$ and $\bf a$ are also separable. It can be divided into $M$ smaller problems related to ${\bf e}^m$ and $a_m$, $m=1,...,M$, respectively. For notational simplicity, denote ${\bm \phi}^m \in {\mathbb C}^{1 \times (M+1 )}$ and ${\bf y}^m \in {\mathbb C}^{1 \times (M+1 )}$ as the $m$th rows of  ${\bm \Phi}$ and $[{\bf P}({\bm \Psi}){\bf W},{\bm \sigma}]$, respectively. Then, the subproblem with respect to ${\bf E}$ and $\bf a$ is an SOCP problem and given by,
\begin{equation}
\begin{aligned}
& \min _{{ {\bf e}^m}, {a_m}}   \Re \{ \text{Tr}\{ ({\bm \phi}^m )^H  {\bf e}^m   \} \}   -  { \zeta_m} {a_m} \\ & \quad\quad\quad\quad\quad\quad+ \tfrac{1}{2\rho}  ( \|  {\bf e}^m - {\bf y}^m  \|_2^2    +  ( { c_m} - { a_m} )^2  ), \\
& \text{s.t.} \quad e_m^m + a_m \ge \sqrt{\gamma_m} \cdot  \| {\bf e}_{-m}^m \|_2,\; \Im\{e_m^m\}=0.
\end{aligned}
\end{equation}
Checking the first-order optimality conditions, we know that the optimal solutions of (21) must satisfy,
\begin{equation}
\left\{\begin{array}{l}
a_m = c_m + \rho ( { \zeta_m} + {\beta_m} ), \\
e_m^m = \Re \{ y_m^m - \rho {\phi}_m^m \} + \rho \beta_m, \\
{\bf y}_{-m}^m - {\bf e}_{-m}^m - \rho {\bm \phi}_{-m}^m \in \rho \beta_m \sqrt{\gamma_m} \cdot \partial \| {\bf e}_{-m}^m \|_2,
\end{array}\right.
\end{equation} 
where $\beta_m$ is the Lagrangian multiplier; ${\phi}_m^m$ and $y_m^m$ are defined similarly as $e_m^m$ and ${\bm \phi}_{-m}^m$; ${\bf y}_{-m}^m$ are defined similarly as ${\bf e}_{-m}^m$; $\partial \| {\bf e}_{-m}^m \|_2^2$ denotes the subgradient of $\| {\bf e}_{-m}^m \|_2^2$, i.e.,
\begin{equation}
\partial \| {\bf e}_{-m}^m \|_2 = \left\{\begin{array}{l}
{\bf e}_{-m}^m / \| {\bf e}_{-m}^m \|_2^2,\; \text{if} \; {\bf e}_{-m}^m \ne 0 \\
 \left\{ {\bf x} \;| \;{\bf x} \in \mathbb{C}^{1 \times M} , \| {\bf x} \|_2 \le 1 \right\},  \; \text{if}\; {\bf e}_{-m}^m = 0
\end{array}\right.
\end{equation} 
Inserting (23) into (22), and applying the KKT conditions, we obtain
the analytical solution of ${\bf e}^m$ as,
\begin{equation}
\begin{aligned}
   & \text{if} \quad  \| {\bf y}_{-m}^m - \rho {\bm \phi}_{-m}^m \|_2 \le \frac{\sqrt{\gamma_m}}{2} [ \Re \{ \rho {\phi}_{m}^m - y_m^m \} - c_m - \rho{ \zeta_m} ]^+ \\
    & \quad e_m^m= \frac{2 \Re \{  y_m^m - \rho {\phi}_{m}^m \}+[ \Re \{ \rho {\phi}_{m}^m - y_m^m \} - c_m - \rho{ \zeta_m} ]^+}{2},\\
    & \quad {\bf e}_{-m}^m = 0, \quad a_m = \frac{2 ( c_m + \rho \zeta_m )+[ \Re \{ \rho {\phi}_{m}^m - y_m^m \} - c_m - \rho{ \zeta_m} ]^+}{2}, \\
    & \text{else}\\
     & \quad \beta_m = \frac{[\sqrt{\gamma_m}\| {\bf y}_{-m}^m - \rho {\bm \phi}_{-m}^m \|_2-\Re \{ { y}_{m}^m - \rho { \phi}_{m}^m  \}-c_m-\rho\zeta_m ]^+}{\rho(2+\sqrt{\gamma_m})}, \\
    & \quad a_m = c_m + \rho ( { \zeta_m} + {\beta_m} ), \quad e_m^m =  \Re \{ y_m^m - \rho {\phi}_{m}^m \} + \rho \beta_m,\\
    & \quad {\bf e}_{-m}^m = ( \| {\bf y}_{-m}^m - \rho {\bm \phi}_{-m}^m \|_2 - \rho \beta_m \sqrt{\gamma_m} ) \frac{{\bf y}_{-m}^m - \rho {\bm \phi}_{-m}^m}{\| {\bf y}_{-m}^m - \rho {\bm \phi}_{-m}^m \|_2}.
\end{aligned}
\end{equation} 

In summary, the BSUM-based approach to solving ${\mathcal P}_{\mathcal{L}{\rho}}$ is outlined in Algorithm 2, which is efficient as each step is computed analytically. The full PDD-based algorithm for problem (6) is obtained by embedding Algorithm 2 into step 2 of Algorithm 1.

\subsubsection{Computation Complexity and Convergence Analysis} The PDD-based algorithm for solving problem (9) has a complexity of approximately ${\cal O}(\text{max}\{M, N, K\}^3)$ per iteration, which is relatively low compared to the problem dimension of $MNK$. Moreover, an important property of Algorithm 1 is that every limit point of the generated sequence is a stationary point of problem (6); see \cite{shi2020penalty} for the detailed proof.

\begin{algorithm}
	\floatname{algorithm}{Algorithm}
	\renewcommand{\algorithmicrequire}{\textbf{Input:}}
	\renewcommand{\algorithmicensure}{\textbf{Output:}}
	\caption{: The BSUM-based algorithm to $\left(\mathcal{P}_{\mathcal{L}{\rho}} \right)$.}
	\begin{algorithmic}[1]
		\REQUIRE 
                 Initialize ${\cal L}_{\rho}^{\text{new}} ({\bf W}, {\bm \theta}, {\bf a},{\bm \Psi}, {\bf E}, {\bf c}; {\bm \Xi}, {\bm \Phi}, {\bm \zeta} )$.\\
            \STATE {\textbf {Reapeat}}
            \STATE ${{\cal L}_{\rho}^{\text{old} }}(\cdot) ={{\cal L}_{\rho}^{\text{new} }}(\cdot)$;
            \STATE  Calculate $\bf W$ by (12) with the bisection method;
            \STATE  Calculate $\bm \theta$ by the closed-form solution in (14) from \cite{9478276};
            \STATE  Calculate ${\bm \psi}_m, \; \forall m$ by (16);
             \STATE  Calculate $\bf c$ by (20);
             \STATE  Calculate ${\bf E}$ and $\bf a$ by (24);
             \STATE  Calculate ${\cal L}_{\rho}^{\text{new}} ({\bf W}, {\bm \theta}, {\bf a},{\bm \Psi}, {\bf E}, {\bf c}; {\bm \Xi}, {\bm \Phi}, {\bm \zeta} )$ by (8);
		\STATE {\textbf {Until } $\tfrac{|{{\cal L}_{\rho}^{\text{old} }}(\cdot)-{{\cal L}_{\rho}^{\text{new} }}(\cdot)|}{{{\cal L}_{\rho}^{\text{old} }}(\cdot)} <\vartheta $.}\\
	\end{algorithmic}%

\end{algorithm}

\section{Numerical Results}
In this section, we compare the proposed method with the following benchmarks: 1) \textbf{AO-SDR} \cite{li2021joint}: solve the joint design problem following the AO algorithm, where its subproblems are solved using the semi-definite relaxation (SDR) method; 2) \textbf{AO-DC} \cite{he2023reconfigurable}: Similar to 1), but the subproblems are solved with rank-one constraints, followed by the difference-of-convex (DC) framework to relax these constraints; 3) \textbf{Without IRS} \cite{matskani2008convex}: solve the AC problem without IRS and; 4) \textbf{Random IRS}: Similar to 3) but with IRS in random phase shift.

In simulations, we use $N=20$, $M=20$, $K=50$, $P = 0$dB, and $\sigma_m=-20$dBm unless otherwise specified. The BS and IRS are positioned at (0, 0) and (50, 10), respectively, with users randomly distributed within a 5-meter radius centered at (70, 0). Channel models are based on \cite{he2023reconfigurable}, with channels represented as the product of large-scale and small-scale fading. Small-scale fading is modeled as a complex zero-mean Gaussian random matrix with unit covariance. Large-scale fading follows the path loss model $\text{PL} = ( \text{PL}_0 - 10 \zeta \log_{10} ( \frac{d}{d_0} ) )$, where \(\text{PL}_0 = -30 \, \text{dB}\), \(d_0 = 1 \, \text{m}\), \(\zeta\) is the path loss exponent, and \(d\) is the link distance. The path loss exponents are 2.2 for the BS-IRS link, 2.5 for both the IRS-user and BS-user links. Results are averaged over 1,000 random fading realizations.

\begin{figure}[!t]
\centering
\includegraphics[width=3.5in]{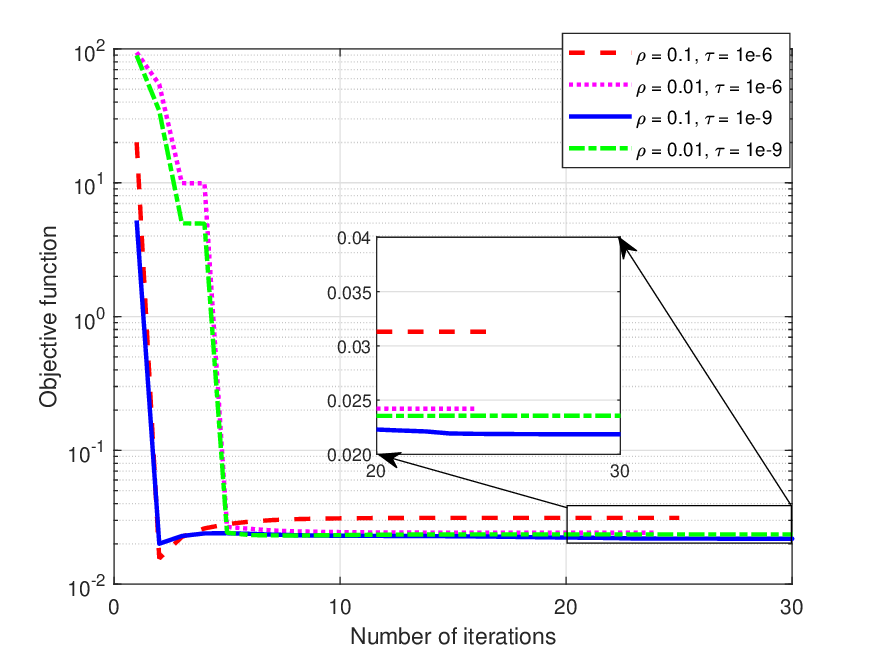}
\caption{Convergence behavior of the PDD-based algorithm.}
\label{fig_1}
\end{figure}

\begin{figure}[!t]
\centering
\includegraphics[width=3.5in]{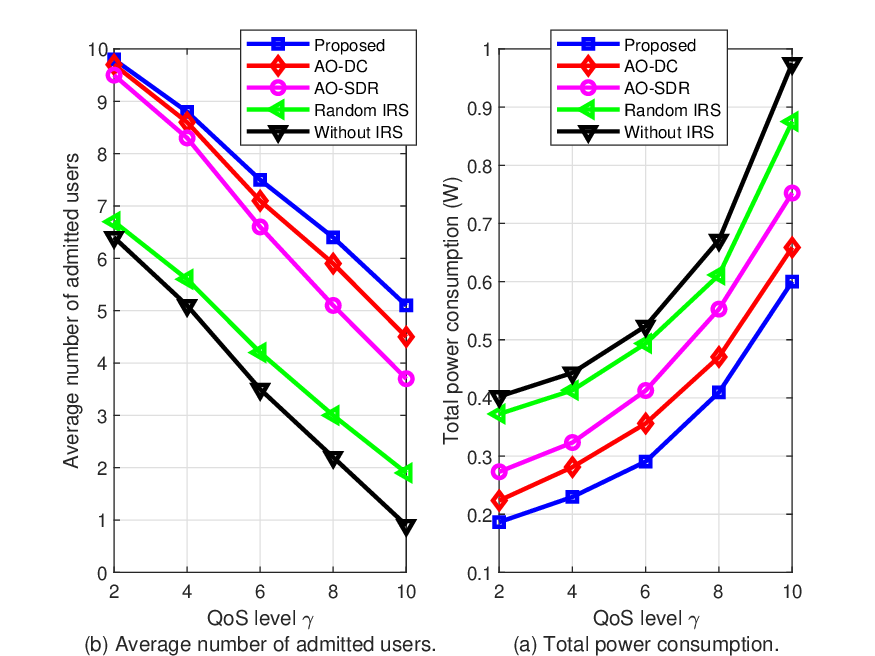}
\caption{Performance comparison versus QoS ($\gamma$).}
\label{fig_1}
\end{figure}

\begin{figure}[!t]
\centering
\includegraphics[width=3.5in]{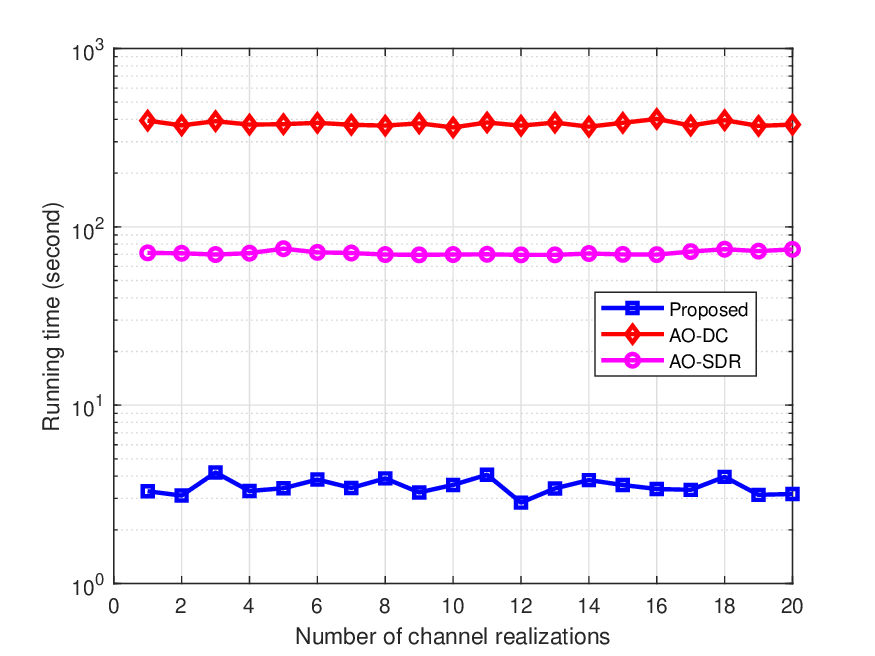}
\caption{Running time comparison between different algorithms.}
\label{fig_1}
\end{figure}

Figure 1 illustrates the convergence of the proposed algorithm. As shown, the algorithm consistently converges to a stable value across different hyperparameters, $\rho$ and $\tau$, demonstrating robust convergence irrespective of hyperparameter selection. This behavior also provides indirect support for the theoretical convergence of the proposed algorithm.

Figure 2 compares the performance of the proposed algorithm with existing methods. Specifically, as shown in Figures 2(a) and 2(b), the proposed algorithm admits 0.4 and 0.9 more users and achieves 0.06W and 0.12W lower power consumption at QoS = 6 compared to AO-DC and AO-SDR, respectively. These results indicate that the proposed algorithm converges to a better stationary point.

Figure 3 compares the computation times of different algorithms by recording the running times for the first 20 randomly generated channel realizations. The proposed algorithm averages 3.5 seconds, about 100x faster than AO-DC (373.5s) and 10x faster than AO-SDR (70.1s). This enhanced efficiency is due to dividing the problem into sub-problems with smaller variable dimensions and using efficient closed-form updates.

\section{Conclusion}
In this letter, we proposed an efficient method for the joint AC and power minimization problem. Specifically, we first approximated the AC indicator using a continuous function and then developed a distributed algorithm based on the penalty dual decomposition (PDD) method to iteratively solve it. The proposed algorithm was highly efficient, achieving stationary convergence with a derived closed-form solution.

\bibliographystyle{IEEEtran}
\bibliography{Bibliography.bib}

\vfill

\end{document}